\documentclass[a4paper]{article}

\makeatletter
\def\@seccntformat#1{\@ifundefined{#1@cntformat}%
   {\csname the#1\endcsname\quad}  
   {\csname #1@cntformat\endcsname}
}
\let\oldappendix\appendix 
\renewcommand\appendix{%
    \oldappendix
    \newcommand{\section@cntformat}{\appendixname~\thesection\quad}
}
\makeatother

\usepackage[dvipdfmx]{graphicx}

\usepackage{cite}
\usepackage[normalem]{ulem}
\usepackage{comment}

\usepackage[dvipdfmx]{xcolor}
\usepackage{amssymb,amsfonts,amsmath}
\usepackage[top=25truemm,bottom=25truemm,left=20truemm,right=20truemm]{geometry}
\usepackage{bm}
\usepackage{cases}
\usepackage{multirow}
\usepackage{slashbox}
\usepackage{array}
\usepackage{subcaption}

\usepackage{lineno}
\newcommand*\patchAmsMathEnvironmentForLineno[1]{
  \expandafter\let\csname old#1\expandafter\endcsname\csname #1\endcsname
  \expandafter\let\csname oldend#1\expandafter\endcsname\csname end#1\endcsname
  \renewenvironment{#1}
     {\linenomath\csname old#1\endcsname}
     {\csname oldend#1\endcsname\endlinenomath}}
\newcommand*\patchBothAmsMathEnvironmentsForLineno[1]{
  \patchAmsMathEnvironmentForLineno{#1}
  \patchAmsMathEnvironmentForLineno{#1*}}
\AtBeginDocument{
\patchBothAmsMathEnvironmentsForLineno{equation}
\patchBothAmsMathEnvironmentsForLineno{align}
\patchBothAmsMathEnvironmentsForLineno{flalign}
\patchBothAmsMathEnvironmentsForLineno{alignat}
\patchBothAmsMathEnvironmentsForLineno{gather}
\patchBothAmsMathEnvironmentsForLineno{multline}
}

\usepackage[ruled,vlined]{algorithm2e}
\usepackage{setspace}	
\SetCommentSty{mycommfont}
\usepackage{multicol} 

\usepackage{setspace}

\usepackage{framed}	
\usepackage{lscape}	
\usepackage{calc}	
\usepackage{soul}	

\usepackage{caption}
\title{Coevolution of social norms and cooperation in public and private situations}

\bigskip
\author{Daiki Miyagawa${}^{1}$, Koki Miyabara${}^{2}$ and Genki Ichinose${}^{2*}$
\ \\
\ \\
${}^{1}$
Graduate School of Science and Technology, Shizuoka University, \\3-5-1 Johoku, Naka-ku, Hamamatsu, 432-8561, Japan\\
${}^{2}$
Department of Mathematical and Systems Engineering, Shizuoka University, \\3-5-1 Johoku, Naka-ku, Hamamatsu, 432-8561, Japan\\
$^*$ Corresponding author (ichinose.genki@shizuoka.ac.jp)}

\begin{document}

\maketitle

\section*{Abstract}

Cooperation in human society is sustained by reputation. In general, the reputation of an individual is determined by others who observe his behavior, but this rarely happens in private situations. This may cause people to behave inconsistently, cooperating in public and not cooperating in private. A previous experiment showed that people gave a lower reputation to an individual who cooperated in public but defected in private rather than a consistently uncooperative individual regardless of public and private situations. However, the reason behind this is unclear. Here, we study how cooperation and the reputational mechanism co-evolve on the condition that two types of interaction (public and private) exist. The simulation results show that the evolved social norm is characterized by at least one of the following: preference for consistent or aversion of inconsistent behavior in both interactions when the risk that behaviors in private interactions are observed exceeds a certain threshold. We also find that such social norms promote cooperation in private situations as well as in public ones.


\section*{Keywords}
Indirect reciprocity,
prisoner's dilemma, 
cooperation,
social norm,
private interaction

\section{Author summary}
Human cooperation is often guided by reputation, where cooperative actions in public are more visible and hence more likely to influence one's reputation. However, this dynamic changes in private settings where actions are less observable. Previous experiments indicated that inconsistent behavior (cooperating publicly but not privately) is viewed less favorably than consistent non-cooperation. This study explores why such reputational mechanisms evolve and how they influence cooperative behavior in different settings.
Using simulations, we demonstrate that social norms evolve to value consistency in behavior across public and private interactions when the likelihood of private actions being observed exceeds a certain threshold. These norms promote cooperation not only in public but also in private situations. We find that these evolved social norms, characterized by either a preference for consistent behavior or an aversion to inconsistency, significantly influence cooperative behaviors in both public and private contexts.

\section{Introduction\label{sec:introduction}}
Cooperation is necessary for society to function.
On the other hand, it is not stable behavior in the sense that it tempts others to engage in exploitation.
To make cooperation stable, there exist several mechanisms \cite{Nowak2006Science}, but the key mechanism among non-relative and large-scale societies is reputation.
Cooperation gives a good reputation to the donor. 
People with good reputations are more likely to be helped by others who observe their good behaviors.
In this way, cooperation is maintained because it creates mutual cooperation among those who act based on reputation, called indirect reciprocity \cite{Alexander1987book, Nowak2005Nature}.
In indirect reciprocity, the rules that determine how reputation is given to the donor are called social norms.
Nowak et al.~show that cooperation can evolve through this indirect reciprocity based on the social norm called image scoring~\cite{Nowak2006Science}.
After this study, various theoretical \cite{Kandori1992RevEconStud, Okuno-Fujiwara1996GamesEconBehav, Nowak1998Nature, Nowak1998JTheorBiol, Leimar2001ProcRoyalSocB, Brandt2004JTheorBiol, Ohtsuki2004JTheorBiol, Panchanathan2004Nature, Ohtsuki2006JTheorBiol, Pacheco2006PLoSComputBiol, Suzuki2007JTheorBiol, Fu2008PhysRevE, Ohtsuki2009Nature} and empirical studies explored many aspects of indirect reciprocity.
Experiments showed that people help those who help others \cite{Wedekind2000Science, Milinski2001royalb, Milinski2002Nature, Milinski2002royalb, Wedekind2002CurrBiol, Bolton2005JPubEcon, Semmann2005BehavEcolSociobiol, Rockenbach2006Nature, Seinen2006EurEconRev, Sommerfeld2007PNAS, Engelmann2009GamesEconBehav, Ule2009Science, Pfeiffer2012JRSocInterface, Molleman2013ProcRSocB}.
Thus, indirect reciprocity is a fundamental basis for promoting cooperation.
Recent representative examples that rely on reputation-based cooperation mechanisms are internet commerce where buyers care about sellers' reputations \cite{Tennie2010TrendsCognSci, Melnik2002JIndustEconom, Bolton2004ManagementSci, Houser2006JEconManagStrategy, Resnick2006ExpEcon, Diekmann2014AmSociolRev, Bolton2013ManageSci, Tadelis2016AnnuRevEcon}.

Although Nowak et al.~\cite{Nowak1998Nature} used a simple model where cooperation (defection) leads to a good (bad) reputation, the impact of such actions on reputation can be influenced by multiple factors, including the reputations of both the donor and the recipient.
In other words, even the same action may give a different reputation depending on who performs the action and who receives it.
When a new reputation is assigned to the donor only from his action, it is called a first-order norm.
When a new reputation is assigned to the donor from both his action and the reputation of the recipient, it is called a second-order norm.
A third-order norm additionally requires the reputation of the donor \cite{Pacheco2006PLoSComputBiol, Santos2018Nature}. 
Ohtsuki et al.~mathematically analyzed social norms used at the third-order level and found the eight robust cooperative social norms \cite{Ohtsuki2004JTheorBiol, Ohtsuki2006JTheorBiol}.
Because there are many other concerns for the concept of indirect reciprocity, it has been investigated from various points of view such as the social norms including the past reputation of actors~\cite{Santos2018Nature}, the reputation of recipients \cite{Murase2023PLoSComputBiol}, social norms to be stochastic \cite{Murase2023PLoSComputBiol}, private assessment of reputations \cite{Fujimoto2022SciRep, Fujimoto2023PNAS}, or the existence of favoritism for those who belong to the same population and prejudice for those who belong to another population~\cite{Whitaker2018SciRep}. 

Many studies of indirect reciprocity assumed that people can observe all interactions and update the reputation of the donor and the recipient based on the observation.  
However, in reality, we often can use only incomplete information, which makes it difficult to build a cooperative society based on indirect reciprocity \cite{Nowak1998JTheorBiol}. 
From this point of view, Ohtsuki et al.~focused on two types of situations: public and private \cite{Ohtsuki2015PLoSComputBiol}.
Public interactions are observed but private ones are not.
Humans are known to change their behavior depending on where they are situated.
For example, people may show good attitudes when they face their boss or colleagues but may show bad attitudes at other times.
Ohtsuki et al.~mathematically showed that some sensitive strategies that change behavior depending on being observed or not are evolutionarily stable as with the ones that are not related to observations.
On the other hand, Kishimoto et al.~experimentally found that whether the donor changed his action with or without observation significantly affected the reputation of observers \cite{Kishimoto2018Psychologia}. 
This evidence suggests that the difference between the situations that others can observe or not have an impact on human decision-making in some aspects. 
However, why humans evolved with such evaluation traits is still unclear.

Otsuki's model showed that given a social norm, the strategy that cooperates in public situations but not cooperates in private situations may be evolutionary stable, but it was not clear why this social norm evolves.
As Kishimoto experimentally showed, whether the behavior is consistent between public and private situations may affect reputation.
Therefore, if the comparison of behavior in both situations is also introduced as a social norm, a strategy of cooperation in public situations, but not in private situations, may evolve.
To verify this hypothesis, we introduce social norms that allow individuals to compare behaviors in public and private games and investigate the co-evolution of social norms and cooperative behaviors through individual-based simulations.
\section{Model}
\begin{figure}[h]	
	\centering		
	\includegraphics
		[width=\columnwidth]	
		{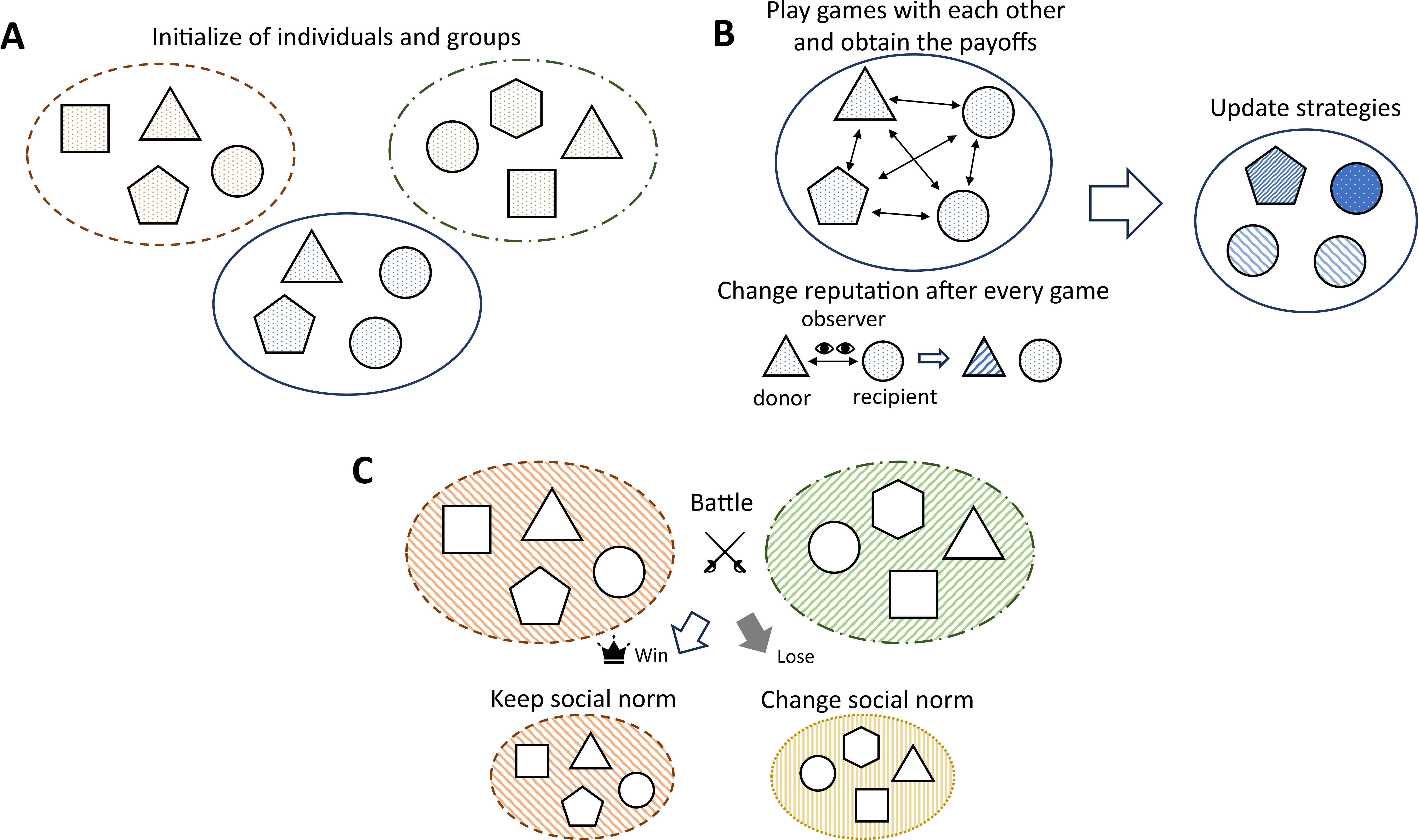}
	\caption{Conceptual figure of the model. {\bf A}: The population consists of many groups (large ellipses) of individuals (circles and polygons). The difference in shapes represents the difference in strategies. {\bf B}: Individuals play a game with others in the same group and obtain the payoffs. The reputations of donors are also changed after every game.
	After all the games are finished, the strategies of individuals are updated. {\bf C}: Social norms shared by individuals in the same group. Social norms give reputations to others. A battle between groups occurs. The norm used in the winning group is taken over by the lost group.}	
	\label{model:image}		
\end{figure}

We develop an individual-based model where social norms and strategies co-evolve.
The conceptual figure of the model is shown in Fig.~\ref{model:image}.
We briefly describe each part here and the details are explained later.
The population consists of many groups of individuals (Fig.~\ref{model:image}A). 
Individuals play a game with others in the same group and obtain the payoffs.
The donor's reputation is changed by observers after every game.
After all the games are finished, the strategies of individuals are updated  (Fig.~\ref{model:image}B). 
Individuals in a group share a single social norm and give a reputation to others depending on the social norm.
Finally, there is a battle between two groups, and the norm used by the winning group is taken over by the losing group (Fig.~\ref{model:image}C).

\subsection{Games and strategies}
First, we describe the game used in the model.
We use donation games as the interaction between individuals. 
The donation game is a two-player game. 
In the donation game, the roles of ``donor'' and ``recipient'' are randomly assigned.
Only the donor takes action on whether to donate according to its strategy.
If the donor chooses to donate, he/she pays $c=1$ as the cost, and the recipient obtains payoff $b=4$.
If the donor chooses not to donate, nothing happens to both payoffs.
Table~\ref{tab:payoff} gives the payoff matrix. The payoff condition satisfies the prisoner's dilemma if $b>c>0$.
Therefore, ``not to donate'' is the dominant strategy. 
Each individual plays the game with every other individual in the group (once for each, hence no direct reciprocity works).

In this study, we consider two types of situations when individuals play the donation games: public and private.  
The difference between them is as follows: the public situations assume that reputation is always public, thus the reputation of a donor is always changed after every game.
In contrast, the private situations assume a private setting where observation rarely occurs with low probability $q\ll1$, and reputation is updated only if observed.
All individuals share their reputation within the group. 
Individuals play the game in public situations with probability $p$ while they play it in private situations with probability $1-p$.



\begin{table}[h]
	\centering
	\caption{Payoff matrix. Every variable denotes the payoff of the ``Self'' player obtained when the two players make their decisions. We use $b=4$ and $c=1$.}
	\begin{tabular}{cccc}
	                      			&               	& \multicolumn{2}{c}{Opponent}		\\ \cline{3-4} 
		                      		&               	& Donate    & Not to Donate   		\\ \hline
		\multirow{2}{*}{Self} 	& Donate        	& $b-c$      & $-c$              	 	\\
	                      			& Not to Donate 	& $b$         & $0$                		\\ \hline
	\end{tabular}
	\label{tab:payoff} 
\end{table}



When an individual plays the donation game, the action is selected according to its strategy.
A strategy is composed of four binary genes (Table \ref{tab:strategy}).
Thus, the number of strategies is $2^4=16$.
The first two genes are used to play the donation games in public situations, and the last two genes are used in private situations.
When a donor takes an action, he compares his reputation with the opponent's reputation.
In the public donation game, if the opponent's reputation is equal to or larger than his reputation, he takes action C (1) or D (0) according to the first locus.
If the opponent's reputation is lower than his reputation, he takes action C (1) or D (0) according to the second locus.
The same criteria are applied to the third and fourth loci for the private donation game.


\begin{table}[h]
	\centering
	\caption{Strategies consist of four binary genes to reflect the two aspects of his/her situation: public or private, and the comparison of reputation between self and opponent.}
	\begin{tabular}{ccccccc}
		Situation               	&\hspace{5mm}& \multicolumn{2}{c}{Public} 			&\hspace{5mm}& \multicolumn{2}{c}{Private} 		\\ \hline
		Reputation relationship 	&\hspace{5mm}& $r_o \geq r_s$	& $r_o < r_s$	&\hspace{5mm}& $r_o \geq r_s$	& $r_o < r_s$	\\
		Action                  	&\hspace{5mm}& C/D          		& C/D         	&\hspace{5mm}& C/D          		& C/D          	\\ \hline
	\end{tabular}
	\label{tab:strategy}
\end{table}





\subsection{Social norm}
A donor's reputation is changed after every game according to the social norm shared in the donor's group. 
Social norms are composed of ten binary genes (Table \ref{tab:norm}).
The first four genes are used in the public situation, and the next four genes are in the private. 
These eight genes take a value $+1$ or $-1$, added to the donor's reputation when the change takes place. 
The last two genes are used to evaluate consistency in the donor's action between public and private situations. 
Those genes change the donor's reputation with $+1$, $0$, or $-1$.
Thus, the number of social norms is $2^8 \times 3^2=2304$.

\begin{table}[h]
	\centering
	\caption{Social norms consist of ten genes separated in two parts. The first eight genes reflect three aspects: public or private, the comparison of reputation between self and opponent, and the donor's action. The last two genes are used to evaluate consistency in the donor's action between the public and private situation. 	}
	\scalebox{0.93}{
		\begin{tabular}{ccccccccccc}
			Situation               	&\hspace{5mm}& \multicolumn{4}{c}{Public}                                     				&\hspace{5mm}& \multicolumn{4}{c}{Private}                                    			\\ \hline
			Reputation relationship 	&\hspace{5mm}& \multicolumn{2}{c}{$r_o \geq r_s$}	& \multicolumn{2}{c}{$r_o < r_s$}	&\hspace{5mm}& \multicolumn{2}{c}{$r_o \geq r_s$}	& \multicolumn{2}{c}{$r_o < r_s$} 	\\
			Donor's action          	&\hspace{5mm}& C			& D        		& C			& D             	&\hspace{5mm}& C              	& D             	& C             	& D             	\\
			Changes                 	&\hspace{5mm}& $ -1/+1$		& $ -1/+1$       	& $ -1/+1$          & $ -1/+1$         	&\hspace{5mm}& $-1/+1$          	& $-1/+1$           & $-1/+1$           & $-1/+1$             	\\ \hline
		\end{tabular}
	}
	\par\vspace{12pt}
	\begin{tabular}{ccc}
		Consistency of actions 	& Achieved 		& Not achieved 	\\ \hline
		Changes                 	& $-1/0/+1$ 	& $-1/0/+1$     	\\ \hline
	\end{tabular}
\label{tab:norm}
\end{table}
Once a game takes place, there are eight cases, depending on which game (public or private), reputation criteria (above or below), and donor's action (C or D).
In the public donation game, the first four loci are used. Then, if the recipient's reputation is equal to or larger than the donor's reputation, the first two loci are used within those four.
The third and fourth loci are used if the recipient's reputation is lower than the opponent's within those four.
Then, the donor's action is checked.
If that action is C, then the first locus is used within those four.
If that is D, then the second locus is used within those four.
The same applies to the third and fourth loci when the recipient's reputation is lower than that of the donor.
The same criteria for the private donation game are applied to the fifth to eighth loci.
In this way, when the reputation is updated, observers add the value in the corresponding locus (one of the eight cases) to the donor's reputation according to the social norm within the group.

\begin{figure}[h]	
	\centering		
	\includegraphics
		[width=0.8\columnwidth]	
		{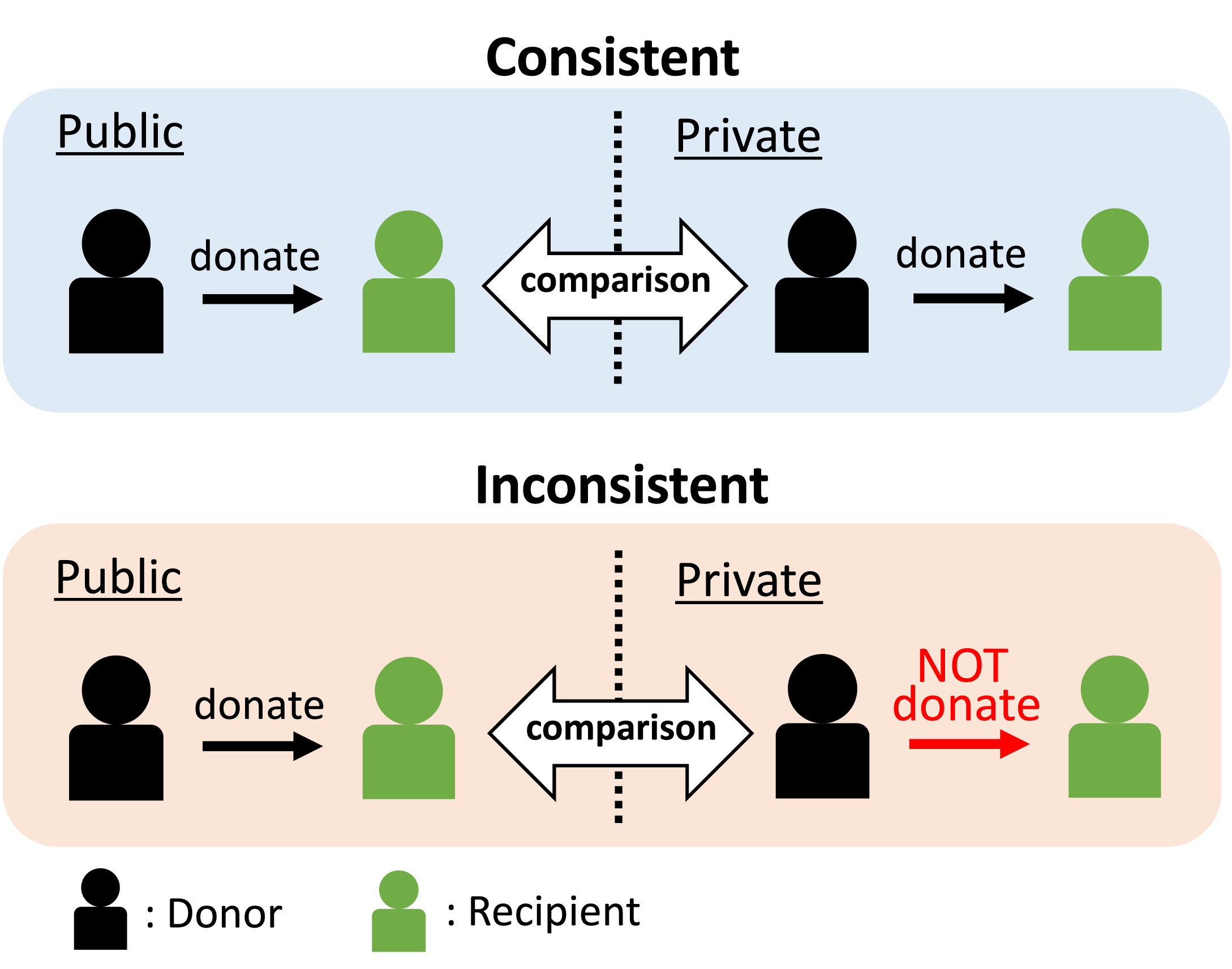}	
	\caption{Comparison of actions in both games.}	
	\label{ActionComparison}		
\end{figure}

Observers also evaluate the donor's actions by comparing actions between the two games.
They do it according to the last two genes determined by the social norm.
Individuals can memorize one action of each donor in each situation (public/private and above/below, in a total of four cases).
Based on this memory, individuals compare the actions between public and private games.
If the donor's action to the same criteria (above or below the donor's reputation) is consistent between the two games (cooperate or defect in both games), the value of the ninth gene is added to the reputation.
If that is inconsistent (cooperate in the public game but defect in the private game or vice versa), the value of the tenth gene is added to the reputation.
An example is shown in Fig.~\ref{ActionComparison}.

Thus, in the reputation change after the game, a two-step reputation change takes place.
Basically, the donor's new reputation is the sum of the donor's current reputation plus the values of the norms corresponding from the first to eight genes, which is determined by an action that is taken in a given situation, and from the ninth and tenth genes, which is determined by the comparison with the other game.
However, if the target individual has not played the public donation game in the past in that generation, if it played the private donation game but was not observed, or if it has no memory for comparison, no evaluation is made by comparison with its behavior in the other game (i.e., the value added to reputation is zero). Also, no changes are made that exceed the range of reputation ($[-5,+5]$), which defines the minimum and maximum.




\subsection{Evolution}
We prepare $M$ groups, each of which has $N$ individuals. 
Every pair of individuals play the donation games in random order. 
After all the games are played, all individuals synchronously update their strategies.
This process is regarded as one generation. 
We repeat it until 10,000 generations. 

The strategy updating is based on the following equation: 
\begin{align}
P_x = \frac{f_x}{\sum_{k=1}^{N} f_k}
\end{align}
which means that individual $i$ selects individual $x$ to imitate his/her strategy in proportion to $f_x$ where $f_x$ is the payoff of individual $x$.
Please note that 1 is added to the donor's payoff in each game to avoid the negative value of $f_x$. 
When an individual copies the strategy of another individual, a mutation occurs for each gene of the strategy with probability $\mu$. Mutation flips the binary value of the gene.



Moreover, two groups are randomly selected after each generation when a conflict between the groups occurs with probability $p_{\rm conflict}=0.3$. 
The two groups compare the average payoffs between groups $\Pi_A$ and $\Pi_B$ where $A$ is the winner (higher payoff) group and $B$ is the loser (lower payoff) group.
As a result, the norm of the group with the lower payoff approaches the norm with the higher payoff.
Specifically, each gene of the norm of the loser group $B$ changes to the corresponding gene of the winner group $A$ with the following probability. 
\begin{align}
P_n = \frac{\Pi_A}{\Pi_A + \Pi_B}
\end{align}
For each gene in the norm, a mutation occurs with probability $\mu_n=0.001$ and the value of the genes flips to the other value (when there are three possible values for a gene, it changes with equal probability to two different values that are not its original value).
After each generation ends and the strategy and norm are updated, all individuals' reputations, payoffs, and memories are initialized.



The evolution of such social norms represents the imitation of other superior groups, wars with other groups, and colonization in the real world \cite{Yu2013ProcediaComputSci}. 
Different groups, such as nations and tribes, have their own specific norms, values, etc., as part of their culture, and these change the probability of survival and group expansion. 
When a war or other competition occurs between groups, the loser group either assimilates into the winner group and adapts the same culture or becomes extinct as the individuals who made up the group disperse. 
There are historical examples of this phenomenon, and the norms that regulate the behavior of economically successful groups are sometimes imitated by neighboring groups \cite{Boyd2009PhilosTransRSocB}.


\section{Result}
In this study, we focus on the probability $p$ that the public games occur and the probability $q$ that the private games are observed, and see how the cooperation rates and the final social norms change when these parameters are varied.
Simulations were performed by varying $p$ from 0.1 to 0.9 with the interval of 0.1, assuming that both public and private games can always occur (i.e., cases, where $p$ is 0 or 1, were not considered).
The maximum value of $q$ was set to 0.10 because the private game was basically unobservable, and $q$ was varied from 0.00 to 0.10 with an interval of 0.01.
The simulations were repeated 100 times for each parameter to use the average values as the results.

\subsection{Typical evolution}
The simulation results roughly showed two typical evolutionary processes.
The first case was that cooperation only evolved in the public donation games but not in the private games.
The other case was that cooperation evolved both in the public and private donation games.
In the next section, we explain thee first case when the ratio of public and private games is equal ($p=0.5$) as the representative parameter.

\subsubsection{First case: Cooperation not evolved in the private games}
\begin{figure}[h]	
	\centering		
	\includegraphics
		[width=\columnwidth]	
		{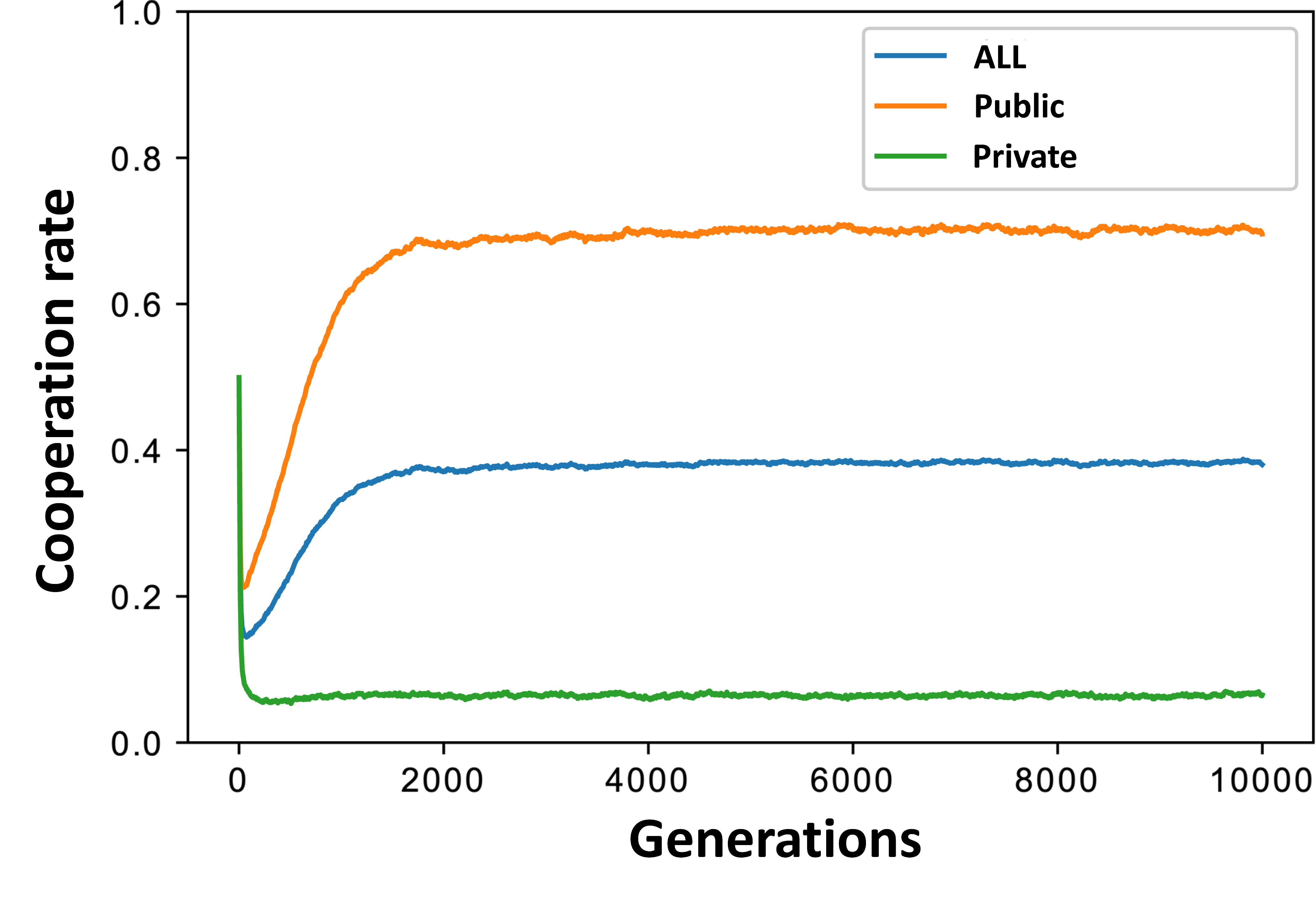}	
	\caption{Change of cooperation rate in the case where cooperation is promoted in public but not in private ($p=0.5,q=0.01$).}	
	\label{fig3-1}		
\end{figure}

As an example of the case in which the cooperation rate increases only in the public games but not in the private games, Fig.~\ref{fig3-1} shows how the cooperation rate changes with the generation (average of 100 trials) when $p=0.5$ and $q=0.01$.
It shows that at the beginning of the generations, the cooperation rate rapidly approaches zero for both public and private games.
This is because only the strategies of individuals evolve by within-group selection.
In contrast, the evolution of social norms by between-group selection does not happen in that early stage of evolution.
Defectors are basically advantageous to cooperators in any group. Thus defectors become dominant first.
However, groups that contain many cooperators can dominate groups that contain many defectors by the rare conflicts among groups because the average payoff of the former groups is higher than that of the latter.
Thus, cooperation gradually spreads among groups.

Nevertheless, the evolution of cooperation by between-group selection occurs only in public games.
In private games, individuals' actions are rarely seen by others, which means that indirect reciprocity does not work.
Thus, the optimal strategy for individuals in this parameter setting is to defect with others within a group because reputation does not matter in private games, while it is to cooperate with others within a group in public games because groups that have social norms based on reputation can dominate other groups which do not have them.
This kind of evolutionary process has been seen in quite low $q$, such as $q=0.01$, with the wide range of $p$, as explained later.

Table \ref{table3-1} shows the evolved norms when $p=0.5$ and $q=0.01$.
It shows that only four games that evaluate the action of public games converge out of ten genes.
The four genes converged to $\{+1, -1, -1, +1\}$.
This means that, in public games, based on the reputation of the donor, cooperating with those who have a high reputation (above the donor) and not cooperating with those who have a low reputation (below the donor) obtains a high reputation ($+1$) and the opposite behavior obtains a low reputation.
This social norm in public games is the same as Stern-judging discovered by Pacheco et al.~\cite{Pacheco2006PLoSComputBiol}.
This result is reasonable because it is known that Stern-judging is a norm that promotes cooperation.

On the other hand, the genes in private games (the next four genes after public games) do not converge.
This is because $q$ is quite small, the action of private games is rarely observed and is not used for evaluation.
Thus, these genes in the norm have little effect on reputation.
In addition, since private games are not observed, the observer cannot compare the action of the donor's public and private games. Therefore, the gene that is evaluated by comparison with another game in the norm (the last two genes) does not converge.

\begin{table}[h]
	\centering
	\caption{Evolved social norm when $p=0.5, q=0.01$. Evaluate as ``Fixed ($+1$ or $-1$)" when the values are the same for at least 90\% of the total population in the last 1000 generations of at least 70\% (70 runs) for each gene, and as ``Not fixed (*)" when they are not.}
	\begin{tabular}{ccccccccccc}
		Situation               	&\hspace{5mm}& \multicolumn{4}{c}{Public}                                     				&\hspace{5mm}& \multicolumn{4}{c}{Private}                                    				\\ \hline
		Reputation relationship 	&\hspace{5mm}& \multicolumn{2}{c}{$r_o \geq r_s$}	& \multicolumn{2}{c}{$r_o < r_s$}	&\hspace{5mm}& \multicolumn{2}{c}{$r_o \geq r_s$}	& \multicolumn{2}{c}{$r_o < r_s$} 	\\
		Donor's action          	&\hspace{5mm}& C		& D        			& C			& D             	&\hspace{5mm}& C              	& D             		& C             	& D             	\\
		Changes                 	&\hspace{5mm}& $+1$		& $-$1       			& $-1$             	& $+1$          	&\hspace{5mm}& *          	& *             		& *             	& *             	\\ \hline
	\end{tabular}
	\par\vspace{12pt}
	\begin{tabular}{ccc}
		Consistency of actions 	& Achieved 		& Not achieved 	\\ \hline
		Changes                 	& * 			& *    		\\ \hline
	\end{tabular}
	\label{table3-1}
\end{table}

\subsubsection{Second case: Cooperation evolved in both games}
\begin{figure}[h]	
	\centering		
	\includegraphics
		[width=\columnwidth]	
		{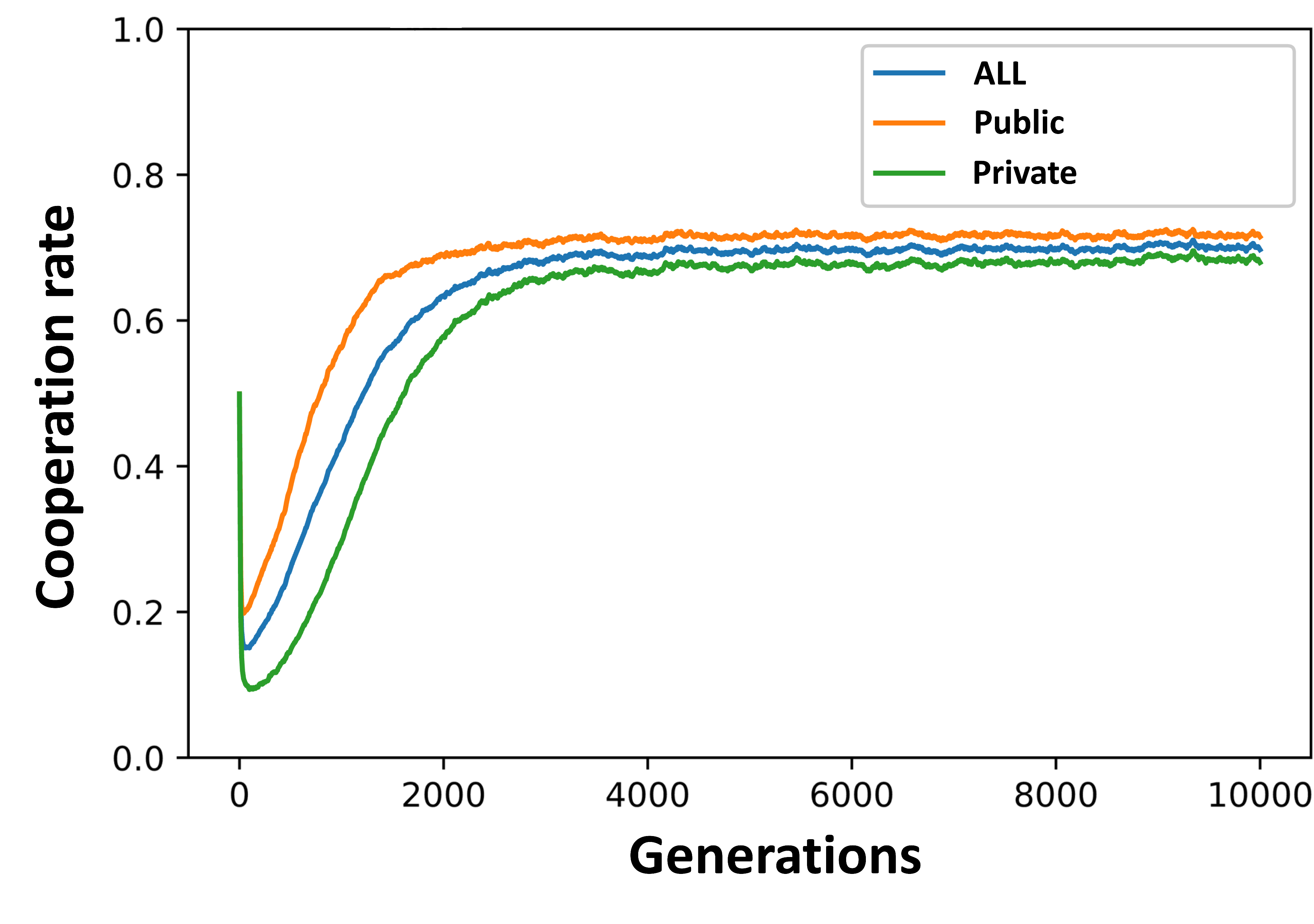}	
	\caption{Change of cooperation rate in the case where cooperation is promoted in both games ($p=0.5,q=0.10$).}	
	\label{fig3-2}		
\end{figure}

Next, as an example of the case in which the cooperation rate increases in both games, Fig.~\ref{fig3-2} shows how the cooperation rate changes with the generation (average of 100 trials) when $p=0.50$ and $q=0.10$.
Even in this case, cooperation rapidly decreases at the beginning of the simulations, as explained above in the case of $p=0.5$ and $q=0.01$.
However, in this case, the fraction of cooperation increases not only in public games but also in private games as the generation progresses.
It means that as $q$ becomes larger, in other words, as private games become more easily observed, the fraction of cooperation increases even in private games.
The larger probability of observing private games suggests that individuals are forced to be cooperative under a norm that promotes cooperation because their behavior affects their reputations.

Table \ref{table3-2} shows the final norm evolved when $p=0.5$ and $q=0.10$.
As for the values of the first four genes that determine the public game's action, they converge to the same form as Stern-judging as in Table \ref{table3-1}.
The next four genes that determine the action in the private game also converge in the form of a high reputation ($+1$) for cooperating with an opponent whose reputation is more than the donor, and a low reputation ($-1$) for not cooperating, which is similar to the evaluation of the action in the public game.

It is noteworthy that the genes (the last two genes) for social norms evaluated by comparison with another game converge to the values of high reputation ($+1$) when ``consistent'' and low reputation ($-1$) when ``inconsistent.''
Because actions are difficult to be observed in private games, reputation rarely changes in those games.
Thus, the advantage of cooperation is small, and defection should be advantageous.
However, when $q$ is larger than a certain value, social norms which favor cooperation evolve.
Under such a situation, social norms that cooperate with opponents with a high reputation and defect with opponents with a low reputation are favored, such as Stern-judging.
Once such a reputation-based society develops, the norms are more sophisticated to evaluate the consistency of actions between public and private games.
In private games, the action of donors is rarely watched, but such sophisticated norms can increase the reputation of the donor when the donor's actions are consistent between public and private games.
At the same time, such sophisticated norms can decrease the reputation of the donor when the donor's actions are inconsistent between public and private games.
Groups with such norms can achieve a high cooperation rate, thus cooperation and norms shown in Table \ref{table3-2} coevolve.

\begin{table}[h]
	\centering
	\caption{Evolved social norm at the final stage of evolution ($p=0.5,q=0.10$).}
	\begin{tabular}{ccccccccccc}
		Situation               	&\hspace{5mm}& \multicolumn{4}{c}{Public}                                     				&\hspace{5mm}& \multicolumn{4}{c}{Private}                                    			\\ \hline
		Reputation relationship 	&\hspace{5mm}& \multicolumn{2}{c}{$r_o \geq r_s$}	& \multicolumn{2}{c}{$r_o < r_s$}	&\hspace{5mm}& \multicolumn{2}{c}{$r_o \geq r_s$}	& \multicolumn{2}{c}{$r_o < r_s$} 	\\
		Donor's action          	&\hspace{5mm}& C		& D        			& C			& D             	&\hspace{5mm}& C              	& D             	& C             	& D             	\\
		Changes                 	&\hspace{5mm}& $+1$		& $-$1       			& $-1$             	& $+1$          	&\hspace{5mm}& $+1$          	& $-1$             	& *             	& *             	\\ \hline
	\end{tabular}
	\par\vspace{12pt}
	\begin{tabular}{ccc}
		Consistency of actions 	& Achieved 		& Not achieved 	\\ \hline
		Changes                 	& $+1$ 		& $-1$     		\\ \hline
	\end{tabular}
	\label{table3-2}
\end{table}

\subsection{Change of cooperation rate depending on $p,\:q$}
\begin{figure}[h]	
	\centering		
	\includegraphics
		[width=\columnwidth]	
		{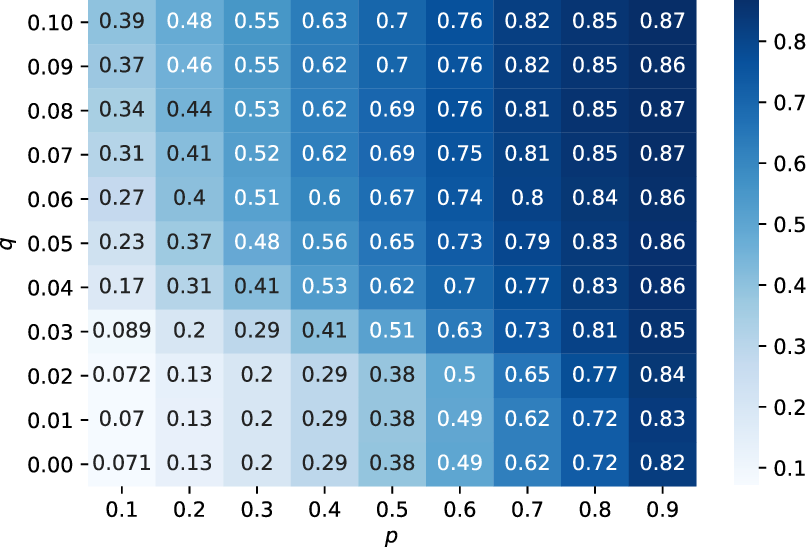}	
	\caption{Average cooperation rate when $p$ and $q$ are changed in both games.}	
	\label{fig3-3}		
\end{figure}

\begin{figure}[h]	
	\centering		
	\includegraphics
		[width=\columnwidth]	
		{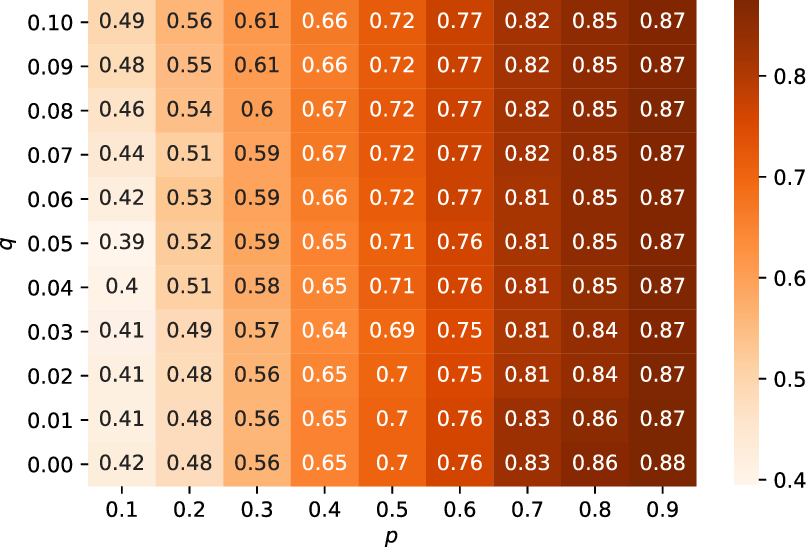}	
	\caption{Average cooperation rate when $p$ and $q$ are changed in public games.}	
	\label{fig3-4}		
\end{figure}

\begin{figure}[h]	
	\centering		
	\includegraphics
		[width=\columnwidth]	
		{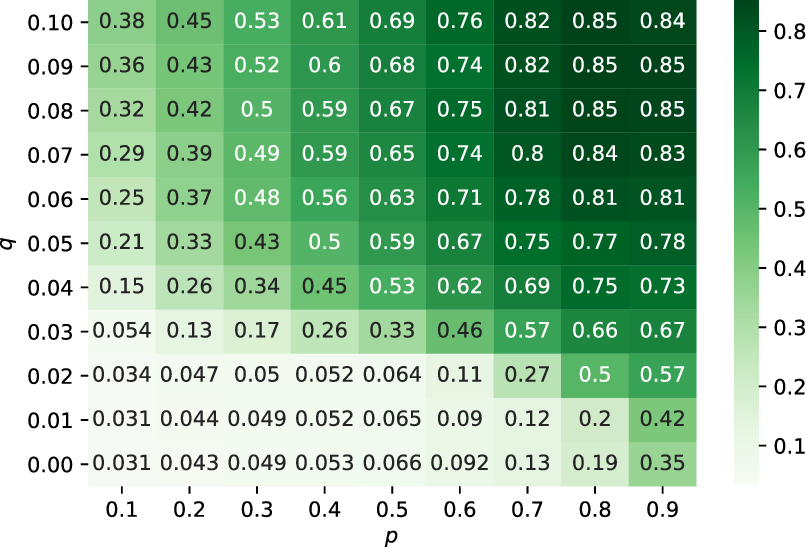}	
	\caption{Average cooperation rate when $p$ and $q$ are changed in private games.}	
	\label{fig3-5}		
\end{figure}

Next, we show the cooperation rate with comprehensive sets of $p,\:q$.
Figures \ref{fig3-3}, \ref{fig3-4}, and \ref{fig3-5} show the average fraction of cooperation in total, public, and private games, respectively, for the last 1000 generations.

The total result (Fig.~\ref{fig3-3}) shows that the cooperation rate becomes larger as $p$ and $q$ increase.
Since $p$ is the fraction of public games and $q$ is the probability that private games are observed, as those parameters increase, the frequency with which individuals are observed becomes large, thus their reputation changes.
This makes reputation more important, and thus indirect reciprocity works strongly to increase the cooperation rate.

Next, we see the cooperation rate in each game.
In public games (Fig.~\ref{fig3-4}), the cooperation rate tends to increase slightly as $q$ gets larger, but this does not affect it much overall.
When $q$ is large, the action in the private game is evaluated more often. Thus the cooperation rate in public games would become large. However, it does not directly affect the public game, so it does not significantly affect the cooperation rate in the public games.
On the other hand, as the ratio of public games $p$ increases, the cooperation rate also increases.
If the frequency of public games is high, the cooperation rate becomes high because observers help the donor when the donor behaves cooperatively and enhance their reputation.
On the other hand, as the probability of private games increases, the likelihood that the opponent helps decreases even if the donor's reputation is high, and thus the merit of behaving cooperatively becomes small. Therefore, when $p$ is small, the cooperation rate is low, even in public games.

The cooperation rate for private games (Fig.~\ref{fig3-5}) is lower than the one for public games because those are rarely observed.
Especially when $q$ is small and almost never observed, the cooperation rate is close to zero.
However, for any value of $p$, once $q$ becomes higher than a certain value, the cooperation rate rapidly increases as $q$ increases, indicating that the cooperation rate tends to be close to that rate in public games.
Also, the cooperation rate increases even in small $q$ when $p$ is large.
The reason for this large change in the cooperation rate is that different parameters lead to a different evolution of norms.
When $q$ is large, norms with specific characteristics can evolve, which can be understood in more detail by looking at the final norm in each parameter.
In the next section, we focus on the evolved social norms by looking at all ranges of $p$ and $q$.

\subsection{Evolved social norms in all ranges of $p$ and $q$}
Here, we present the final evolved norms for the entire $p$ and $q$ ranges. To explain specific genes, we add a symbol to each locus from $l_{1}$ to $l_{10}$ as shown in the following Table~\ref{table3-3}.

\begin{table}[h]
	\centering
	\caption{Symbols assigned to each locus in the norm.}
	\begin{tabular}{ccccccccccc}
		Situation               	&\hspace{5mm}& \multicolumn{4}{c}{Public}                                     				&\hspace{5mm}& \multicolumn{4}{c}{Private}                                    			\\ \hline
		Reputation relationship 	&\hspace{5mm}& \multicolumn{2}{c}{$r_o \geq r_s$}	& \multicolumn{2}{c}{$r_o < r_s$}	&\hspace{5mm}& \multicolumn{2}{c}{$r_o \geq r_s$}	& \multicolumn{2}{c}{$r_o < r_s$} 	\\
		Donor's action          	&\hspace{5mm}& C		& D        			& C			& D             	&\hspace{5mm}& C              	& D             	& C             	& D             	\\
		Changes                 	&\hspace{5mm}& $l_{1}$		& $l_{2}$    			& $l_{3}$             	& $l_{4}$          	&\hspace{5mm}& $l_{5}$         	& $l_{6}$             	& $l_{7}$             	& $l_{8}$             	\\ \hline
	\end{tabular}
	\par\vspace{12pt}
	\begin{tabular}{ccc}
		Consistency of actions 	& Achieved 		& Not achieved 	\\ \hline
		Changes                 	& $l_{9}$ 		& $l_{10}$     		\\ \hline
	\end{tabular}
	\label{table3-3}
\end{table}

\subsubsection{Social norms evolved in public games}
\begin{figure}[h]	
	\centering		
	\includegraphics
		[width=\columnwidth]	
		{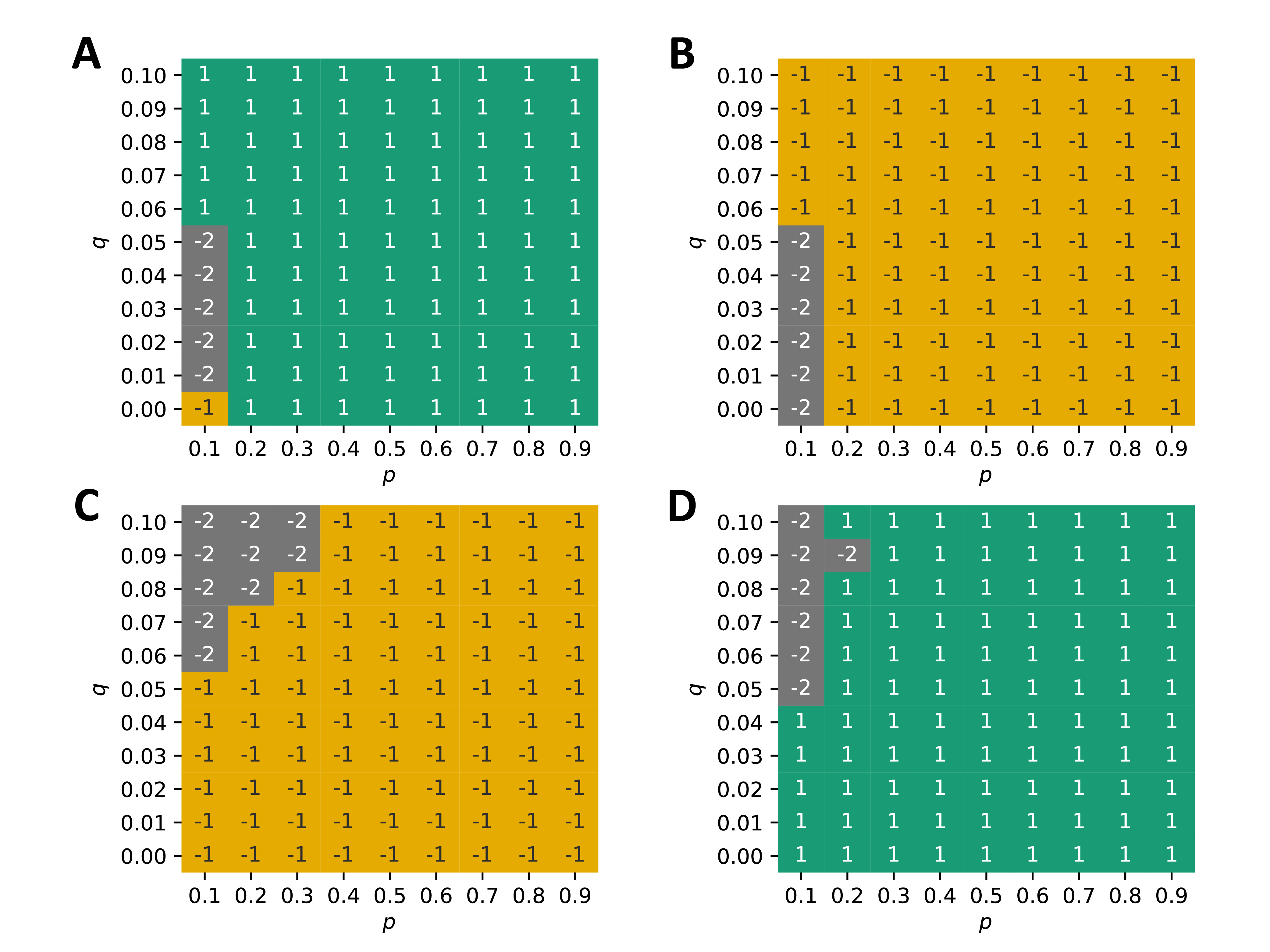}
	\caption{Evolved social norms in each of {\bf A}: $l_1$, {\bf B}: $l_2$, {\bf C}: $l_3$, and {\bf D}: $l_4$ when $p$ and $q$ are varied.}	
	\label{fig3-6}		
\end{figure}

First, we see the evolved social norms in public games.
The genes $l_{1}$ to $l_{4}$ evaluate the action in public games.
Figure \ref{fig3-6} shows the final values of $l_{1}$ to $l_{4}$ in all ranges of $p$ and $q$.
In that figure, $-2$ represents the situation that the value does not converge.
In the wide $p$ and $q$ ranges, the convergence to $l_{1}=1, l_{2}=-1, l_{3}=-1, l_{4}=1$ are observed.
Thus, in public games, as seen in the typical evolution (Sec.~3.1), the norm evolves to a Kandori (Stern-judging).
However, around $p=0.1$ (on the left side of each figure), we see that the norms do not converge.
When most games are private games, even under a norm that promotes cooperation such as Stern-judging, there will be fewer situations in which cooperation is advantageous, making it difficult for cooperative groups to evolve.
As a result, the evolution of the norms is likely to be unstable.

\subsubsection{Social norms evolved in private games}
\begin{figure}[h]	
	\centering		
	\includegraphics
		[width=\columnwidth]	
		{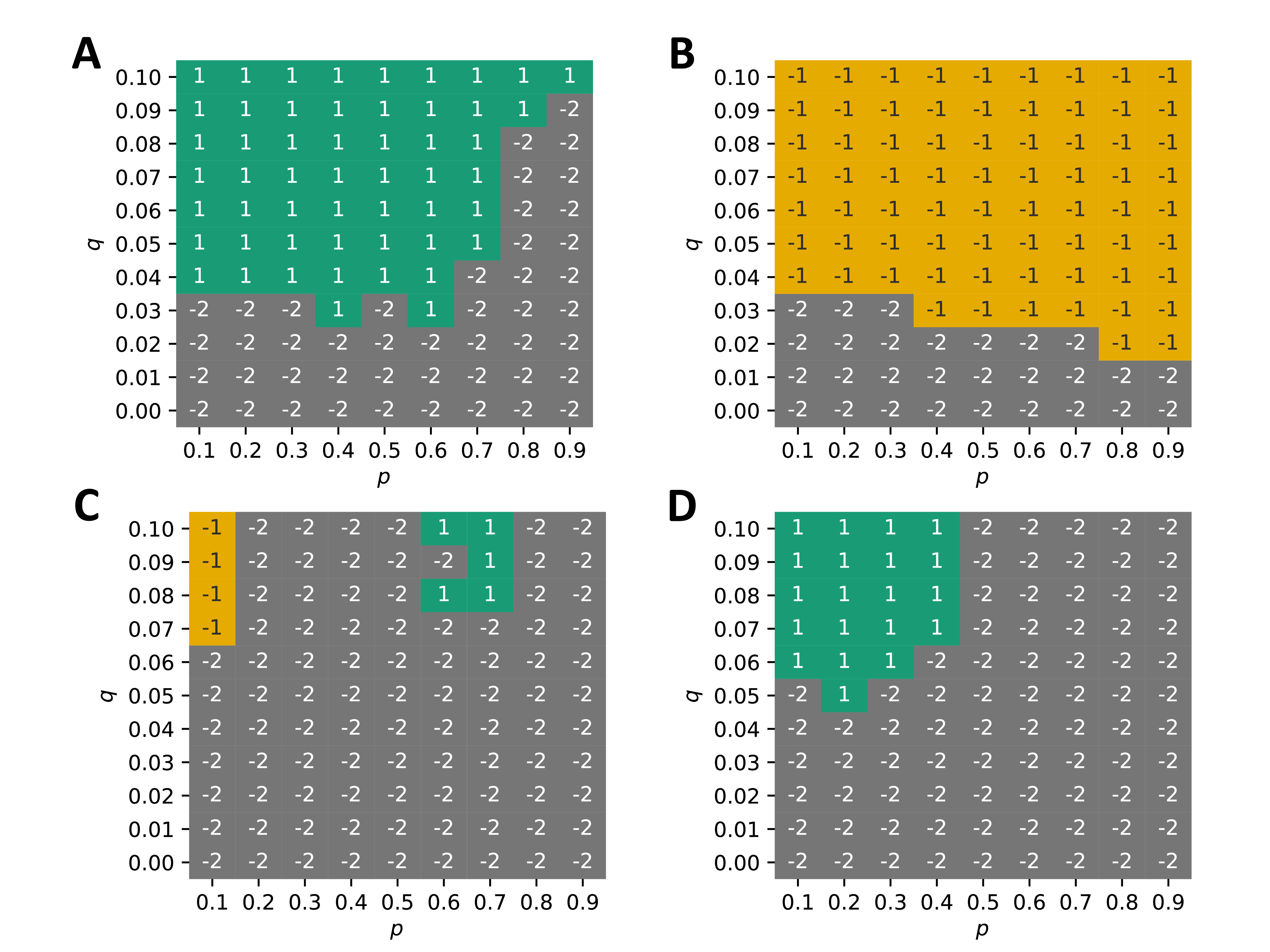}
	\caption{Evolved social norms in each of {\bf A}: $l_5$, {\bf B}: $l_6$, {\bf C}: $l_7$, and {\bf D}: $l_8$ when $p$ and $q$ are varied.}	
	\label{fig3-7}		
\end{figure}

Next, we see the evolved social norms in private games.
The genes $l_{5}$ to $l_{8}$ evaluate the action in private games.
Figure \ref{fig3-7} shows the final values of $l_{5}$ to $l_{8}$ in all ranges of $p$ and $q$.
When $q$ is too small, there is no convergence for $l_{5}$ to $l_{8}$ genes because the private game is almost never observed and individuals are uncooperative in any norm.
However, $l_{5}, l_{6}$ genes almost converge when $q$ is greater than about $0.4$, and when they do converge, as in the norm that promotes cooperation in the previous study, cooperating with a partner with a high reputation (more than a donor) is highly evaluated ($l_{5}=1$) and not cooperating is lowly evaluated ($l_{6}=-1$).
Also, $l_{7},$ and $l_{8}$ genes may not converge even when $q$ is above a certain value, or they may converge differently from the public game.
This may be because, as \cite{Ohtsuki2006JTheorBiol} showed, $l_{5}$ and $l_{6}$ are constant ($1,-1$), while $l_{7}$ and $l_{8}$ can promote cooperation in a variety of ways.
The reason why the evaluation of actions toward a partner with a high reputation ($l_{5}, l_{6}$) is constant ($1,-1$) but the evaluation of actions toward a partner with a low reputation ($l_{7}, l_{8}$) is not is thought to be because $l_{7}, l_{8}$ can promote cooperation in various ways.

\subsubsection{Evolved social norms for behavioral consistency in public and private games}
\begin{figure}[h]	
	\centering		
	\includegraphics
		[width=\columnwidth]	
		{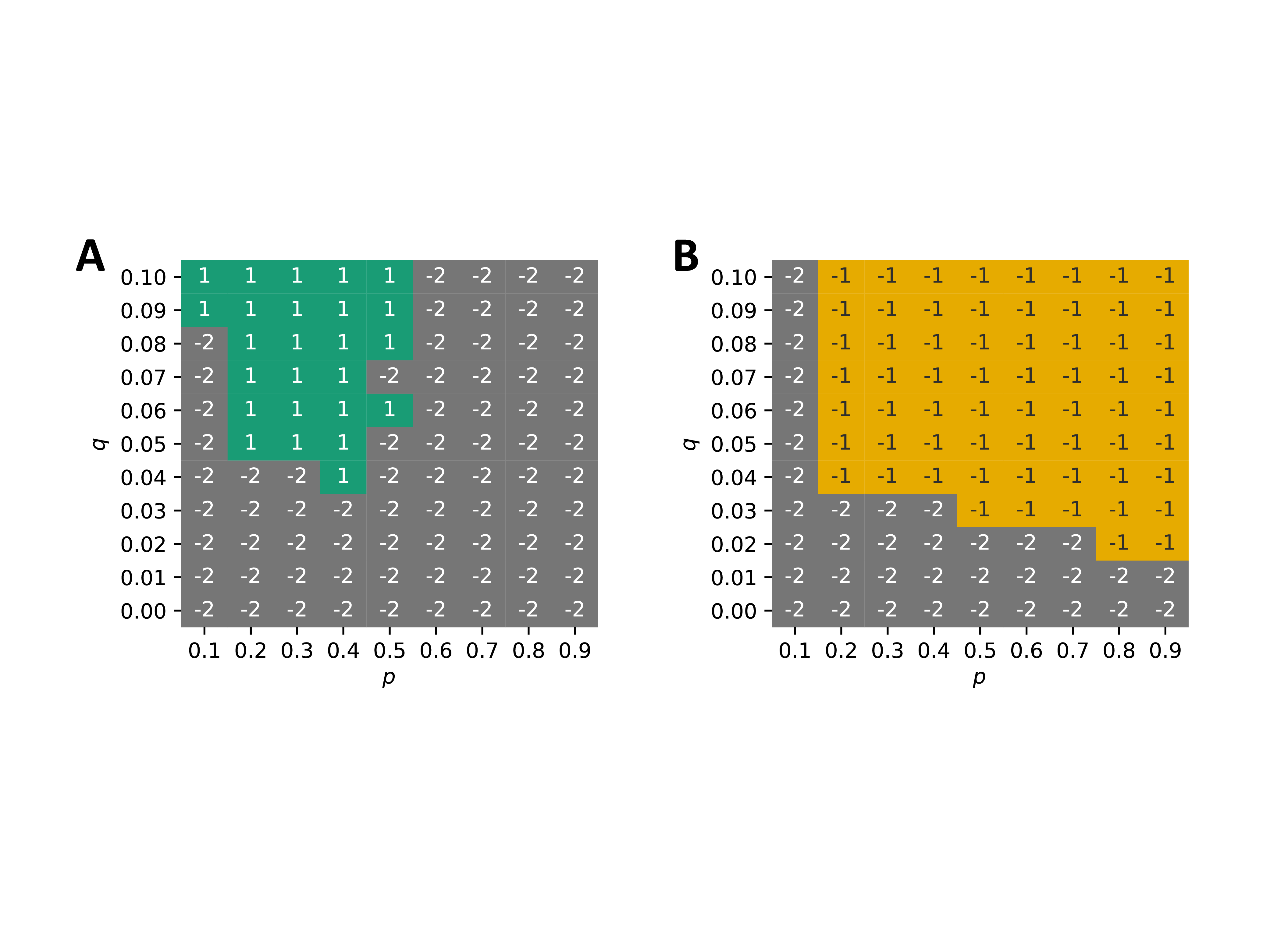}
	\caption{Evolved social norms in each of {\bf A}: $l_9$ and {\bf B}: $l_10$ when $p$ and $q$ are varied.}	
	\label{fig3-8}		
\end{figure}

Figure \ref{fig3-8} show the final values of $l_{9}$ and $l_{10}$ genes over the entire $p, q$ range.
These genes evaluate the actions that were consistent ($l_9$) and not ($l_{10}$) between the two games.
The figure shows that both $l_{9}, l_{10}$ genes tend to converge when $q$ is large, and the larger $p$ is, the more they converge even at smaller $q$.
In the case that convergence happens, $l_{9}$ (reputation for consistency) is almost always high ($1$), and $l_{10}$ (reputation for inconsistency) is almost always low ($-1$).
In other words, when both converge, those who do not change their behavior depending on whether it is public or private will have a relatively high reputation in the group.
As described before, such a norm can create a difference in reputation between individuals who behave cooperatively only in public games and those who behave cooperatively in both games, forcing individuals to be cooperative in both.
Therefore, unlike other norms, it is possible to increase the frequency of cooperation and evolve not only in public games but also in private games.
However, if $q$ is sufficiently small and private games are often not observed, then it is still advantageous for individuals not to cooperate in private games because they can reduce the cost they pay.
That is, when $q$ is small, there is an advantage in reducing the cost of cooperation by cooperating only in private games.
On the other hand, if $q$ is large, there is a disadvantage that reputation continues to be deducted in public games by non-cooperative behavior being observed in private games.
Therefore, depending on the size of $q$, a norm is likely to evolve in which consistent behavior is valued highly (inconsistent behavior is valued lowly).
Also, focusing on $p$, if $p$ is large, the number of private games is small, the merit not to cooperate becomes low, but conversely, in public games, where the number of games is large, the disadvantage of having one's reputation reduced by inconsistent behavior becomes large.
Therefore, the larger $p$ is, the more likely individuals are to cooperate in a norm where inconsistent behavior is valued lowly, and the more likely $l_{10}$ is to converge to $-1$ even at lower $q$.
Convergence of $l_{9}$ to $1$ has the same meaning as $l_{10}$ being $-1$ in the sense that it makes the reputation of a consistent individual relatively high, but since reputation does not become infinitely high, the reputation of an individual with consistent behavior due to $l_{9}$ being $1$ does not necessarily make the reputation of an individual with consistent behavior higher each time. Even if $1$ is added, it does not necessarily mean that the individual with consistent behavior will have a higher reputation than the individual with inconsistent behavior. Therefore, it is considered to be less likely to converge than $l_{10}$.

\begin{figure}[h]	
	\centering		
	\includegraphics
		[width=\columnwidth]	
		{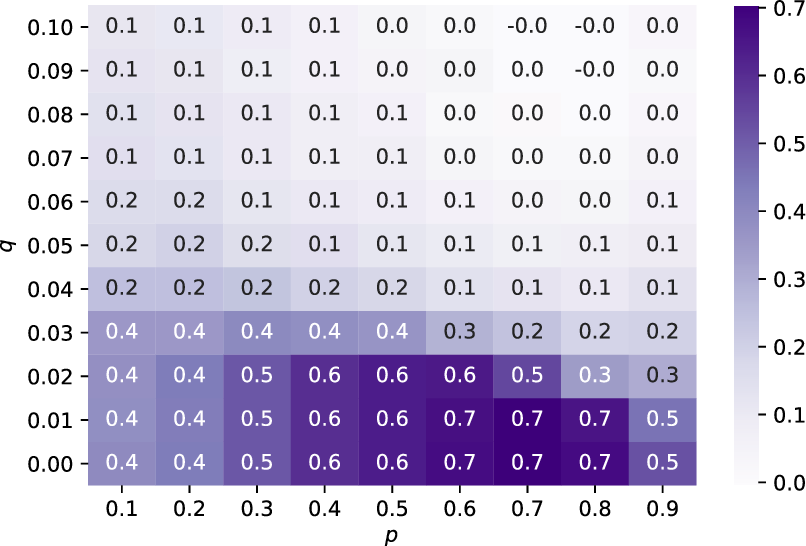}	
	\caption{Difference between the cooperation rates of public and private games for each $p,\:q$ (``cooperation rate of public games'' $-$ ``cooperation rate of private games'').}	
	\label{fig3-16}		
\end{figure}

Figure \ref{fig3-16} shows that, consistent with the above discussion, the difference between public and private cooperation rates almost disappears in the range of $p, q$ such that either the norm's ``rating when consistent ($l_{9}$)'' converges to a high rating ($1$) or the ``rating when inconsistent ($l_{10}$)'' converges to a low rating ($-1$) or at least one of them, indicating that individuals are evolving to cooperate at the same level in the two games.
Conversely, to the extent that there is no such convergence, the cooperation rate for private games is much lower than for public games, indicating that individuals are evolving to be noncooperative only in private games.
In particular, focusing on $q$, there is almost no difference in cooperation rates between public and private situations in the range where $q$ is generally greater than 0.05.
Therefore, if private situations are observed to some extent (albeit with a small probability), the norm will at least, either favor consistent behavior or not favor inconsistent behavior.
Then, individuals in the population evolve cooperatively in both situations.

\section{Discussion}

In this study, we investigated how social norms and cooperation evolve by comparing the behaviors of an individual in two situations to reputation criteria in a model of indirect reciprocity, in which two situations, public and private, exist and an individual can decide whether to cooperate with another individual depending on the situation and both of their reputations.

The results show that in public games, as in previous studies, social norms evolve that positively value the behavior of cooperating with partners with high reputations and not cooperating with partners with low reputations, and individuals evolve using such cooperative behavior.
On the other hand, in private games, social norms are less likely to evolve (converge) because the games are less likely to be observed, and when the probability of observation is high, a social norm that promotes cooperation evolves as in public games, in which cooperation with those with high reputations is rated high and not to cooperate with them is rated low.

An important result is that if private games are observed with more than a certain degree of probability, social norms evolve that favor consistent or disfavor inconsistent behavior in public and private situations, and under such social norms, individuals evolve to behave cooperatively in both situations.
This result indicates that, contrary to the intuitive idea that it is advantageous not to cooperate in private situations that are less likely to be observed and less likely to affect reputation, it may be advantageous to cooperate in both situations to preserve high reputations, even if only a few private situations are observed.


\section*{Acknowledgment}
This study was partly supported by JSPS KAKENHI, Grant Number JP22H01713 (G.I.).

\section*{Additional Information}
Declarations of interest: none. 


\end{document}